\documentclass[twocolumn]{emulateapj}
\usepackage{multirow}
\usepackage{graphicx}
\usepackage{footnote}
\usepackage{natbib}
\usepackage[fleqn]{amsmath}
\usepackage{amsmath}
\usepackage{color}

\shorttitle{Optimized radio follow-up strategy for rare Ib/c SNe}
\shortauthors{Carbone \& Corsi}

\begin{document}

\title{An optimized radio follow-up strategy for stripped-envelope core-collapse supernovae}

\author{Dario Carbone\altaffilmark{1,$\star$}, Alessandra Corsi\altaffilmark{1}}
\altaffiltext{1}{Department of Physics and Astronomy, Texas Tech University, Box 1051, Lubbock, TX 79409-1051, USA}
\altaffiltext{$\star$}{Email: \email{dario.carbone@ttu.edu}}

\begin{abstract}
 
Several on-going or planned synoptic optical surveys are offering or will soon be offering an unprecedented opportunity for discovering
larger samples of the rarest types of stripped-envelope core-collapse supernovae (SNe), such as those associated with relativistic jets,
mildly-relativistic ejecta, or strong interaction with the circumstellar medium (CSM). Observations at radio wavelengths are a useful tool
to probe the fastest moving ejecta, as well as denser circumstellar environments, and can thus help us identify the rarest type of
core-collapse explosions. Here, we discuss how to set up an efficient radio follow-up program to detect and correctly identify radio-emitting
stripped-envelope core-collapse explosions.
We use a method similar to the one described in \citealt{Carbone2018}, and determine the optimal timing of GHz radio observations assuming a
sensitivity comparable to that of the Karl G. Jansky Very Large Array. The optimization is done so as to ensure that the collected radio
observations can identify the type of explosion powering the radio counterpart by using the smallest possible amount of telescope time.
We also present a previously unpublished upper-limit on the late-time radio emission from supernova iPTF\,17cw.
Finally, we conclude by discussing implications for follow-up in the X-rays.
\end{abstract}

\keywords{}

\section{Introduction} \label{sec:intro}
How exactly massive stars die is still an open question as the zoo of supernova (SN) explosions is rather variegate
\citep{Filippenko1997,GalYam2017}.
The most extreme and rare type of core-collapse, stripped-envelope SNe are the engine-driven ones associated with relativistic ejecta
(gamma-ray bursts; GRBs).
The link between broad-lined (BL) SNe of type Ib/c (i.e. stripped of their hydrogen and possibly helium envelopes) and long GRBs has been
established long ago, with the first clear association being that of SN\,1998bw and GRB\,980425 \citep{Galama1998,Kulkarni1998}.
While this link is solid \citep{Woosley2006}, it remains unclear what makes some BL-Ic SNe launch ultra-relativistic jets (GRBs).
In fact, while most SNe associated with GRBs are of type BL Ic, not all BL-Ic SNe are associated with GRBs
\citep[e.g.,][]{Berger2003, Soderberg2006c, Corsi2016}.
A notable example of a relativistic BL-Ic SN without a detected GRB is SN\,2009bb \citep{Soderberg2010}, which showed a fast-evolving
radio counterpart but no high-energy emission. Sources like this
might represent a population of events with properties in between that of ordinary BL-Ic SNe (without ultra-relativistic jets) and GRBs
(hosting ultra-relativistic jets).

The $\gamma$-ray energy of most GRBs with a spectroscopic SN association is
lower than that of cosmological GRBs \citep{Amati2002,Mazzali2014}, suggesting that these GRBs 
may represent a distinct population of intrinsically lower-energy events \citep{Bromberg2011,Waxman2004}, or ordinary GRBs observed off-axis
\citep{Yamazaki2003,Eichler1999}.
Although to date no unambiguous discovery of an off-axis long GRB has been reported, off-axis events are a natural expectation of the
standard fireball model \citep[e.g.][]{Rhoads1997,Piran2004}.
An off-axis GRB jet harbored within a SN explosion may only become visible at late times in the radio
\citep{Perna1998,Waxman2004}, and could more generally represent a potential source of radio emission with peak timescales of about
10-100\,d since explosion, depending on the GRB kinetic energy and obsever's viewing angle.

In order to understand the link between low-luminosity GRBs and engine-driven SNe, larger statistical samples of BL-Ic  SNe with
radio observations are needed.  In the past decade, population studies of BL-Ic SNe have been limited by the rarity of these events
\citep[see e.g.][]{Soderberg2006b}. 
Recently, we have begun to make progress toward carrying out a systematic study of BL-Ic SNe in the radio
\citep{Corsi2011,Corsi2014,Corsi2016}, thanks to the much-increased rate of BL-Ic discoveries enabled by the
intermediate Palomar Transient Factory
\citep[iPTF;][]{Law2009}, and its successor, the Zwicky Transient Facility
\citep[ZTF;][]{Smith2014, Bellm2016, Ho2019}.
Future transient surveys such as the Large Synoptic Survey Telescope \citep[LSST;][]{LSST2009}
will dramatically increase the number of BL-Ic SNe discoveries in optical \citep[$\sim\,10^4$ per year;][]{Shivvers2017}.
It is thus reasonable to expect that several of these sources could be followed-up (and possibly detected) in the radio, allowing to more
stringently constrain the fraction of radio-bright BL-Ic SNe related to long GRBs \citep{Corsi2016}.

Stripped-envelope core-collapse SNe with non-relativistic ejecta but  interacting strongly with dense circumstellar medium (CSM),
can also be accompanied by radio-loud emission and, with limited follow-up observations, may be confused with off-axis GRBs
\citep{Corsi2014, Palliyaguru2019, Salas2013}.
Indeed, because of the lower ejecta speeds, radio emission from strongly interacting SNe tends to peak at later times 
\citep[generally speaking, for a given radio luminosity, the later the peak time, the smaller the ejecta speed; see e.g. ][]{Berger2003}.
The bright radio emission from the recently-discovered and much celebrated AT\,2018cow may also have a CSM origin
\citep[][]{Prentice2018,Margutti2019}.

In light of the above considerations, in this paper we build a methodology for setting up an efficient radio follow-up program aimed at detecting
and correctly identifying radio-emitting stripped-envelope core-collapse explosions with the Karl G. Jansky Very Large Array (VLA).
Our paper is organized as follows. In Section \ref{sec:new_obs} we present new observational results for the engine-driven SN iPTF\,17cw,
which will be used in our analysis. In Section~\ref{sec:methods} we summarize our simulation method; in Section~\ref{sec:results} we discuss
our results for an optimized radio follow-up strategy; in Section~\ref{sec:xrays} we elaborate on the detectability of accompanying emission
in the X-rays; finally, in Section~\ref{sec:conclusions} we conclude.

\section{New VLA observation of iPTF\,17cw}\label{sec:new_obs}
We observed the field of the BL-Ic SN iPTF\,17cw \citep[][]{Corsi2017} with the VLA under our program VLA/18A-240 (PI: Corsi)
on 2018 April 19 UT (at an epoch of about 467 days since iPTF\,17cw optical discovery), when the array was in its A configuration.
This observation was carried out in both C-band (nominal central frequency of $\approx 5$\,GHz), and S-band (nominal central frequency
of $\approx 3$\,GHz). We used J0920+4441 as our phase calibrator, 3C48 as flux and bandpass calibrator.
VLA data were reduced and calibrated using the VLA automated calibration pipeline which runs in the Common Astronomy
Software Applications package \citep[CASA;][]{McMullin2007}. When necessary, additional flags were applied after visual inspection
of the data. Images of the observed field were formed using the CLEAN algorithm (Hogbom 1974), which we ran in interactive mode.
iPTF\,17cw is not detected in the formed images down to a 3$\sigma$ limit of 10~$\mu$Jy in C-band,
and 15~$\mu$Jy in S-band.

We have incorporated the 5\,GHz upper limit derived here in the radio light curve of iPTF\,17cw  used for this study (see Section \ref{sec:models}
for further discussion).
This data point is highlighted with a red star in Figure~\ref{fig:models}.
The late-time non-detection  of iPTF17cw is compatible with expectations that its radio light curve followed a temporal behavior similar
to that of SN\,1998bw \citep[see Fig. 9 in][]{Corsi2017}.

\section{Methods}\label{sec:methods}
\begin{figure*}
\centering
\includegraphics[scale=0.7,viewport = 0 0 445 320, clip]{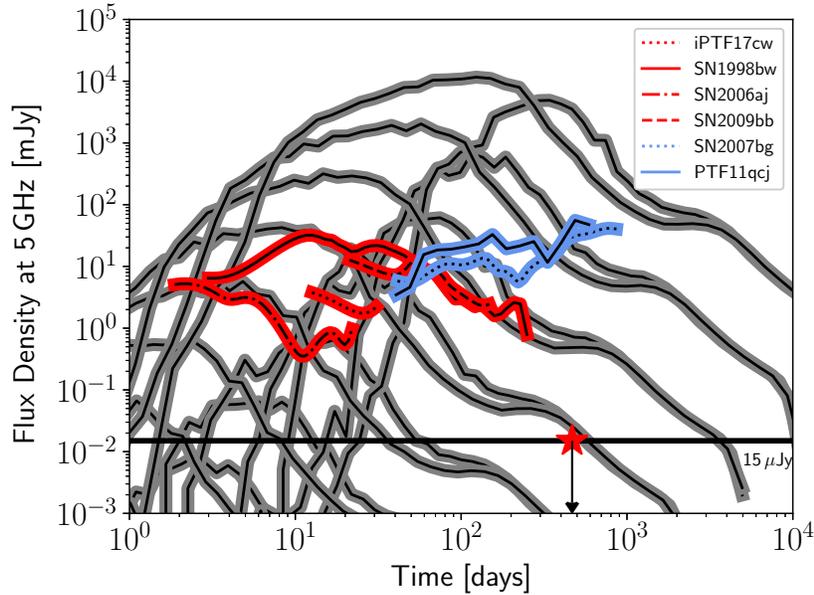}
\caption{
Template and model 5\,GHz light curves of relativistic SNe (red), CSM-interacting SNe (light blue), and off-axis GRBs (grey).
All light curves are scaled to  $z=0.01$.
The red star represents the upper limit derived from the new VLA observation of iPTF17cw presented in Section~\ref{sec:new_obs}.
}
\label{fig:models}
\end{figure*}

To establish the optimal observational strategy for detecting and correctly identifying the nature of radio-bright BL-Ic SNe
(relativistic explosion, off-axis GRB, or CSM-interacting event), we use a simulation method similar to the one presented in
\citealt{Carbone2018}.
We study the optimal observational strategy as a function of the type of explosion we aim to target (Figure~\ref{fig:models};
see Section~\ref{sec:models} for more details).

Thus, we perform our simulations in three steps.
First we adopt relativistic SNe as targets, and use CSM-interacting SNe and off-axis GRBs as ``contaminants''.
The last are radio counterparts that may all be found in association with stripped-envelope core-collapse SNe,
and that we want our radio follow-up observations to distinguish from the radio light curve of our targets with the minimum possible
number of observations.
Next, we treat CSM-interacting SNe as targets, and relativistic SNe and off-axis GRBs as contaminants.
Finally, we adopt off-axis GRBs as targets, and relativistic SNe and CSM-interacting SNe as contaminants.

For each target we assume a known position and distance since radio observations are assumed to follow optical identification, and the
last is likely to provide accurate localization, host galaxy identification, and redshift measurement. We simulate our targets to have
$0.01 \lesssim z \lesssim  0.1$. The largest redshift we choose is motivated by the fact that it corresponds to the largest distance
at which state-of-the-art optical telescopes can observe SNe Ibc, which have an absolute peak magnitude in $r$-band typically between
$\approx -18$\,mag and $\approx -19$\,mag
\citep[for comparison, ZTF has a 5$\sigma$ detection limit of $\lesssim 20.5$\,mag in $r$-band;][]{Graham2019}.
We simulate three different redshift scenarios: one where all target sources have $z=0.01$, a second where all sources have $z=0.1$,
and a third where sources are located at redshifts randomly selected between these boundaries.

As we describe in more detail in what follows, we simulate 10000 realizations of each target (Table~\ref{tab:models}) 
by randomizing the time of the first radio observation ($t_{\rm radio,0}$). We rescale and interpolate the fluxes from our templates where
needed in order to match the times and redshifts we simulate. We then determine the minimum number of radio follow-up observations
(and their corresponding epochs) required to maximize
the probability of correctly and uniquely associating the observed fluxes with those expected from the correct target,  when the observed
fluxes are compared with our bank of radio light curves (including contaminants; see Table~\ref{tab:models}).
We set a maximum of ten on the total number of radio observations that can be performed for each target.
This is a reasonable assumption for a typical one-semester time allocation on the VLA, considering that each epoch in our simulations consists
of a 2\,hr-long observation.

\begin{table}
\begin{center}
\caption{
Summary of light curve templates and models used in this work. Data reported here are taken from:
\citet{Foley2006} and \citet{Kulkarni1998} for SN\,1998bw;
\citet{Strauss1992} anf \citet{Soderberg2010} for SN\,2009bb;
\citet{Sollerman2006} and \citet{Soderberg2006} for SN\,2006aj;
\citet{Corsi2017} for SN\,iPTF17cw;
\citet{SDSS_IV2006} and \citet{Corsi2014} for PTF11qcj;
\citet{Salas2013} for SN\,2007bg.
}
\label{tab:models}
\begin{tabular}{lllll}
\hline
\hline
      &Type &	Redshift	\\
\hline
SN\,1998bw & Rel SN & 0.00867	\\ 
SN\,2009bb & Rel SN & 0.0099		\\ 
SN\,2006aj & RelSN & 0.0331		\\ 
iPTF\,2017cw & Rel SN & 0.093	\\ 
\hline
PTF11qcj & CSM-Int SN & 0.02811	\\ 
SN\,2007bg & CSM-Int SN & 0.0346	\\ 
\hline
E48\_theta45 & LGRB & -	\\
E49\_theta45 & LGRB & -	\\
E50\_theta45 & LGRB & -	\\
E51\_theta45 & LGRB & -	\\
E52\_theta45 & LGRB & -	\\
E53\_theta45 & LGRB & -	\\
E54\_theta45 & LGRB & -	\\
\hline
E48\_theta24 & LGRB & -	\\
E49\_theta24 & LGRB & -	\\
E50\_theta24 & LGRB & -	\\
E51\_theta24 & LGRB & -	\\
E52\_theta24 & LGRB & -	\\
E53\_theta24 & LGRB & -	\\
E54\_theta24 & LGRB & -	\\
\hline
\end{tabular}
\end{center}
\end{table}

\subsection{Radio light curve models and templates} \label{sec:models}
In order to simulate our targets, we use template radio light curves derived from real radio observations of relativistic
and CSM-interacting SNe (rescaled at the simulated redshifts and interpolated to the simulated observation time),
as well as models for off-axis GRB radio light curves.

For relativistic and CSM-interacting SNe, template and simulated fluxes at specific epochs are derived by performing a linear
interpolation between the two closest available data points. If the simulated observing time falls after the time range covered by actual
observations of the source, we perform a linear extrapolation of the flux using the last two available observations.
This can potentially lead to major errors in the simulated flux if the simulated observing time is far from the last epoch of the actual observations,
but we note that this does not happen in the optimized strategy we report in Section~\ref{sec:results}.
Finally, we treat simulated observations at epochs earlier than the first actual observation of a given source in two different ways.
First, we assume that these observations result is non detections (i.e., we assume that the simulated flux falls below our sensitivity;
see Section~\ref{sec:observations}). Next, given uncertainties in the early-time rising behavior of radio SN light curves, we perform
early-time extrapolations by assuming a temporal behavior that mimics that of relativistic SNe for which earlier-time observations
are actually available (see Section \ref{sec:early}). Results from these two different methods for treating early-time epochs are
compared and contrasted to explicitly assess the importance of early-time detections.

To build a set of relativistic SN templates as described above, we use the light curves of SN\,1998bw \citep{Kulkarni1998},
SN\,2006aj \citep{Soderberg2006}, SN\,2009bb \citep{Soderberg2010}, and iPTF\,17cw \citep{Corsi2017}.
We choose the first two because they are GRB-associated SNe with well-sampled radio light curves, SN\,2009bb because it may
represent an event in between GRBs and engine-driven SNe, and iPTF2017cw because it is a relativistic SN located much farther away
than the others.
SN\,1998bw and SN\,2009bb were very bright, nearby events visible from few days up to hundreds of days
after explosion. They were observed in several radio bands, and here we focus on their detectability at
5\,GHz. SN\,2006aj was much fainter, it was detected early on, and decayed rapidly until it became too dim after 20 days since explosion.
It was observed in several radio bands, but the most complete radio light curve is the one at 8\,GHz.
For consistency, in our simulations we extrapolated the 8\,GHz light curve of SN\,2006aj to 5\,GHz using a spectral index $\beta$=0.7,
typical of optically thin synchrotron emitting sources \citep[e.g.,][]{Kellermann1964}, where we use the convention $F_{\nu}\propto \nu^{-\beta}$.
Actual spectral index measurements of SN\,2006aj are compatible with this value within large uncertainties \citep{Soderberg2006}.
iPTF\,17cw was detected 12 days after explosion, and its 5\,GHz light curve also decayed rapidly, with the source becoming undetectable
after 30 days.
Template radio light curves of relativistic SNe at 5\,GHz are plotted in red in Figure~\ref{fig:models}.

Similarly, we use the measured light curves of PTF\,11qcj \citep{Corsi2014} and SN\,2007bg \citep{Salas2013} as templates for our
CSM-interacting SNe.
We note that AT\,2018cow may be an interesting member of this class of SNe \citep{Margutti2019}.
Unfortunately, at the time of writing only a small number of GHz radio detections have been published for this event, so we do not include it here.
Template radio light curves of CSM-interacting SNe at 5\,GHz are plotted in light blue in Figure~\ref{fig:models}.

Finally, we simulate 5\,GHz radio light curves of off-axis high- and low-luminosity long GRBs using BOXFIT v2 \citep{vanEerten2012}.
The BOXFIT light curves depend on several parameters: the luminosity distance ($d_L$); the jet half-opening angle ($\theta_j$);
the viewing angle ($\theta_v$); the total explosion energy  ($E_{\rm iso}$); the interstellar medium density ($n_{\rm ISM}$);
the power-law index of the shocked electrons' energy distribution $(p)$; and the fraction of the energy converted into magnetic fields and
elections ($\epsilon_B$ and $\epsilon_E$). Here we set $\epsilon_B=10^{-2}$, $\epsilon_E=10^{-1}$, $\theta_j$=12\,deg, $p$=2.5
\citep{Ghirlanda2005,Goldstein2016,Beniamini2017}, and $n_{\rm ISM}=1$\,cm$^{-3}$
\citep[which is the average value found in long GRBs;][]{Chandra2012,Granot2014}.
We create models both using $\theta_v=45$\,deg, and $\theta_v=24$\,deg, and varied $E_{\rm iso}$ between $10^{48}$ and $10^{54}$\,erg 
\citep{Frail2001,Ghirlanda2004,Amati2006,Nava2012,Perley2014,Goldstein2016}.
These parameters cover both cosmological, more energetic GRBs, and low luminosity ones that are more common in our cosmic
neighborhood.
The resulting light curves are plotted in grey in Figure~\ref{fig:models} (where we neglect redshift corrections).

\subsection{Monte Carlo simulations} \label{sec:observations}
For each of the targets, we generate 10000 observed light curves drawing from Gaussian distributions with mean equal
to the model/template flux at each epoch, and standard deviation equal to the quadrature sum of the error in the template/model and the flux error
affecting the simulated observations, $\sqrt{\sigma^2_{\rm model/template}+\sigma^2_{\rm obs}}$. In the above, $\sigma_{\rm model/template}$
is set to the interpolated measurement errors for our template SN light curves, and to a nominal 10\% of the flux for off-axis GRB models;
$\sigma_{\rm obs}$ is set to $5\,\mu$Jy, comparable to the image RMS achievable at 5\,GHz with the JVLA in its most compact configuration
(which provides a conservative estimate of the sensitivity) for a total observing time (including overhead) of 2\,hrs per epoch, with $\approx$ 15\%
bandwidth loss on a nominal 4\,GHz bandwidth (3 bit) due to RFI.
At any epoch when the simulated model flux is below our detection threshold of $3\times$RMS, the measured flux is set to
zero and the error on it is set equal to the noise RMS.

\subsection{Optimizing the radio follow-up campaign}\label{sec:delays}
We assume that the first radio observation is always carried out as soon as possible, at
$t_{\rm radio,0}=t_{\rm opt,0}+\Delta T_0$ (having assumed $t=0$ as the time of the SN explosion).
Here $t_{\rm opt,0}$ accounts for delay between the SN explosion and the  optical discovery.
We randomize $t_{\rm opt,0}$ uniformly in the range $1\,{\rm h}-30$\,d.
$\Delta T_0$ allows for a possible further delay between the optical discovery and the earliest radio observation.
We tested three different ranges: $\Delta T_0=1\,{\rm hr}-2$\,d (hereafter dubbed high-urgency follow up),
$\Delta T_0=3-5$\,d (hereafter referred to as medium-urgency follow up), and $\Delta T_0=7-15$\,d (low urgency).

The ultimate goal of our simulations is to determine the minimum number of radio follow-up observations, $n_{\rm min}$
(where 1$\le n\le$10), and their corresponding epochs $\Delta T_n=t_{\rm n}-t_{\rm radio,0}=M_n\times 2$\,d
(where $M_n$ is an integer in the range $1\le M_n \le 183$, and its maximum value of 183 is chosen so that all observations happen within
one year), required to maximize a figure of merit which we refer to as the number of unique and correct associations, computed as follows.

For each of the simulated observations of target light curves, we determine which templates/models
(both targets and contaminants) predict fluxes that at $t_{\rm radio,0}$ agree with the simulated observation within
$3\times\sqrt{\sigma^2_{\rm model/template}+\sigma^2_{\rm obs}}$.
These models/templates are considered positive associations for the first
epoch, and carried forward to the next observing epoch.

The second epoch can happen with any time delay, $\Delta t_2=M_2\times 2$\,d where $M_2=1,2,3,...$, with respect to the
first observation. In general, only a subset of the models that represented positive associations for epoch one will also be positive
associations for epoch two (i.e. will show agreement between observed flux and predicted model/template flux at that epoch within
$3\times\sqrt{\sigma^2_{\rm model}+\sigma^2_{\rm obs}}$). Thus, we optimize the value of $M_2$ by maximizing the number
of associations that in epoch two become unique (only one model/template fits the observed target in both epochs) and correct (the model/template
that fits the observations uniquely is also the correct one, i.e. it is the same template/model from which the observations were simulated). 

We then add a third observing epoch, keeping the two already analyzed in place. As before, the third epoch can happen
with any time delay, $\Delta t_3=M_3\times 2$\,d where $M_3=1,2,3,...$, with respect to the first observation. So we optimize
$M_3$ by maximizing the number of associations that, after being positive in both epoch one and two, become unique and
correct identifications in epoch three.

We keep repeating this process until we reach a maximum of ten epochs. Naturally, adding more observational epochs will
progressively increase the fraction of unique and correct associations up to that epoch.
Note that it may happen that in the optimization process $M_n$ turns out to be larger than $M_{n+1}$. Thus, the times
of the optimized observational epochs are ordered in increasing delays since first epoch once the optimization process is
completed, as presented in Tables~\ref{tab:relSN}, \ref{tab:relSN_early}, and \ref{tab:CSMSN}.

\section{Results}\label{sec:results}

\subsection{Discovering relativistic SNe} \label{sec:relSN}
Our goal here is to optimize the observational strategy to detect relativistic, engine-driven SNe in the radio. For this reason,
we treat the relativistic SN templates listed in Table \ref{tab:models} as our targets, while CSM-interacting SNe and off-axis GRBs
are treated as contaminants.
Targets that are not detectable because always too faint at the considered redshift are excluded from this analysis (i.e., they are
not considered when calculating the efficiency of our strategies)
A summary of our results is reported in Table~\ref{tab:relSN}.
We find that five observations are required in order to maximize the number of unique and correct associations.
In terms of urgency, we find the largest number of unique and correct associations when adopting a high-urgency strategy, i.e., minimum interval
between the optical discovery and the first radio observation (see Section \ref{sec:delays}).

At $z=0.01$, 95\% of the simulated targets are detectable, and 97\% of the detectable targets are uniquely and correctly associated.
All targets simulated from the templates of SN\,1998bw, SN\,2009bb and iPTF\,17cw, and 80\% of SN\,2006aj are detectable.
100\% of the detectable targets simulated from the templates of SN\,1998bw, SN\,2009bb and iPTF\,17cw, and 84\% of SN\,2006aj
are uniquely and correctly associated.
Missed associations are due the fact that SN\,2006aj is very faint and faded away quickly.

For sources at $z=0.1$,
we have much fewer unique and correct associations, especially for targets whose peak flux was close to our detection limit at $z=0.01$ 
(e.g., SN\,2006aj-like sources). In fact, the fraction of detected sources drops to 72\%, and 78\% of those sources is uniquely and
correctly associated.
In particular, 100\% of the targets simulated
from the templates of SN\,1998bw and SN\,2009bb,  19\% of SN\,2006aj, and 71\% of iPTF\,17cw are detectable.
Of those, 100\% of the targets simulated from the templates of SN\,1998bw and SN\,2009bb, 0\% of SN\,2006aj,
and 36\% of iPTF\,17cw are uniquely and correctly associated.
For sources at $z=0.1$ the difference between different urgency strategies is negligible.
This is likely due to the fact that at this distance the early, rising part of the light curve falls below our detection threshold,
thus having early observations would not particularly affect the results. 
Missed associations in this case are due the fact that the light curve iPTF\,17cw is not well sampled at early times.
Our discussion in Section~\ref{sec:early} clarifies this point.

In the case of sources with redshift randomly distributed between 0.01 and 0.1 our results are, as expected, in between the two
previously described cases.
Specifically, 92\% of all simulated sources are detected, and 87\% of the detected sources are uniquely and correctly associated.
In particular, 100\% of the targets simulated from the templates of SN\,1998bw and SN\,2009bb, 
69\% of SN\,2006aj, and 99\% of iPTF\,17cw are detected.
Of those detected sources, 100\% of the targets simulated from the templates of SN\,1998bw and SN\,2009bb, 
48\% of SN\,2006aj, and 88\% of iPTF\,17cw are uniquely and correctly associated.

\begin{figure*}
\centering
\includegraphics[scale=0.5,viewport = 0 0 445 335, clip]{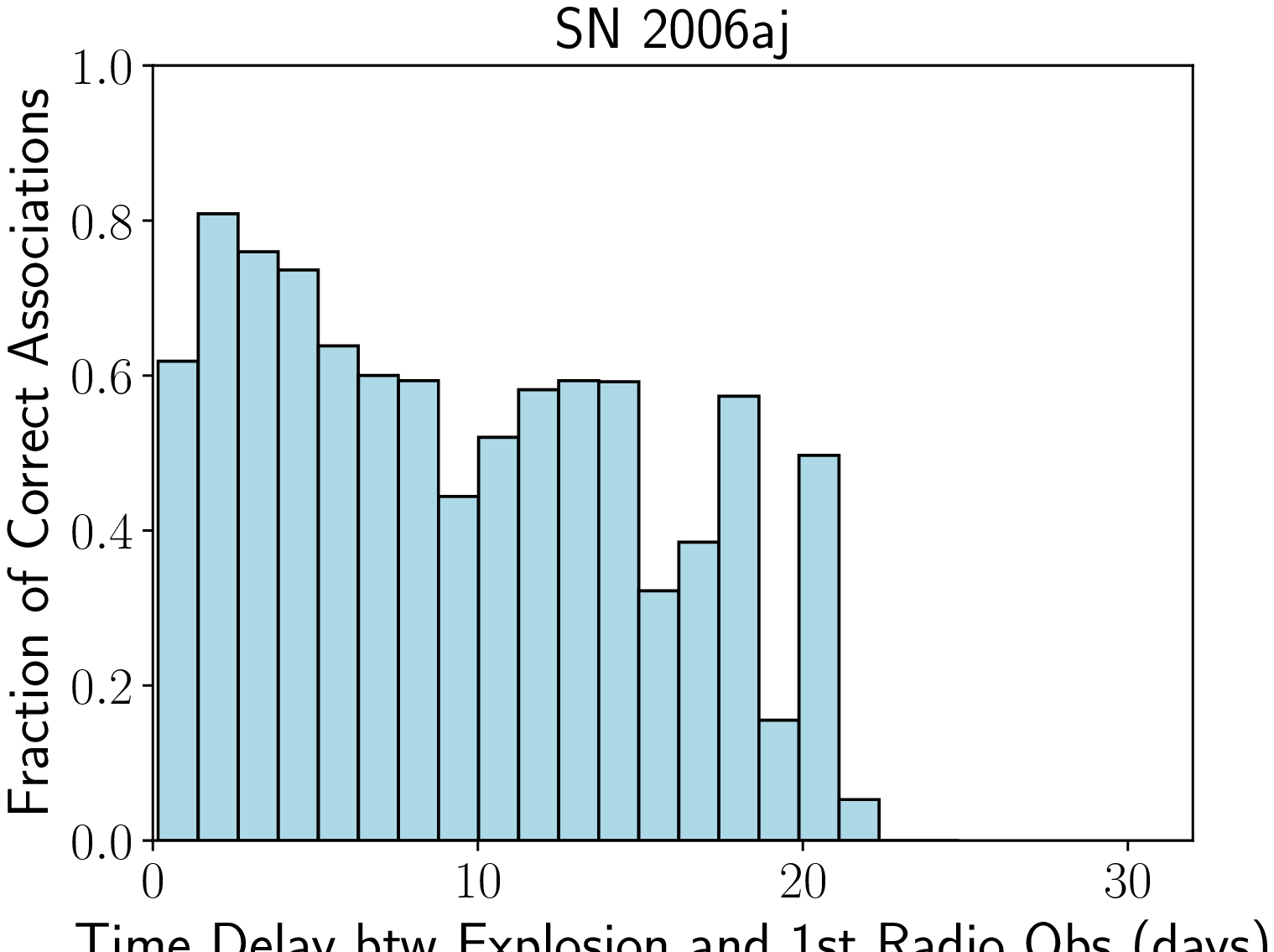}
\includegraphics[scale=0.5,viewport = 0 0 445 335, clip]{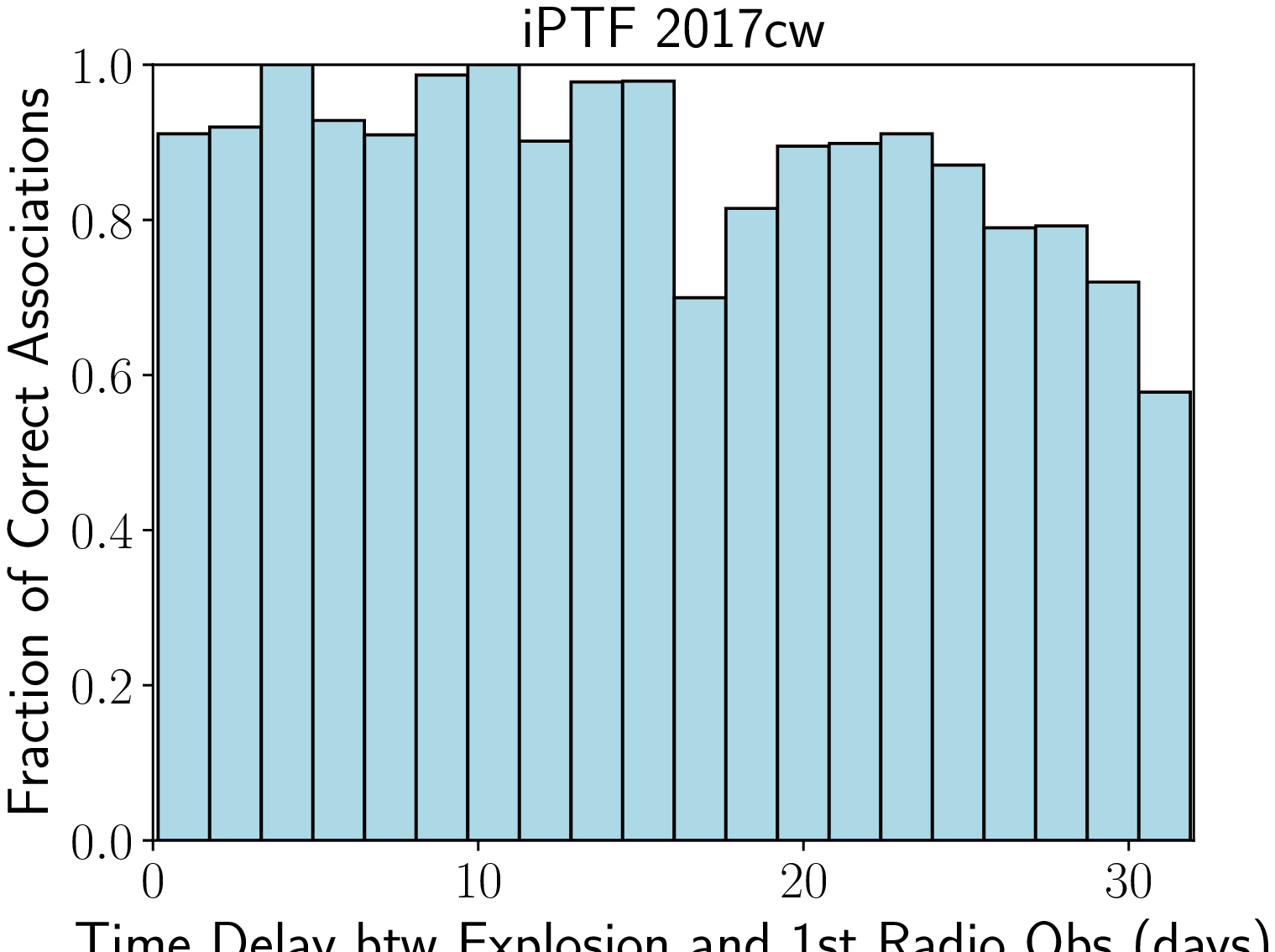}
\caption{
Fraction of targets that are uniquely and correctly associated as a function of the total delay between the SN explosion and
the first radio observation for sources with $z=0.01-0.1$.
}
\label{fig:histo}
\end{figure*}

For completeness, in Figure~\ref{fig:histo} we also plot the the efficiency of the optimized observing campaign as a function of total time
delay for SN\,2006aj-like and PTF17cw-like targets, with redshifts randomly distributed between 0.01 and 0.1.
The total time delay is calculated as the sum of the delay between the explosion and the optical detection, and the delay between the
optical detection and the first radio observation (which defines the urgency of the radio observing campiagn; see Section \ref{sec:delays}).
In this Figure the efficiency (fraction of unique and correct associations) is calculated relative to the number of simulated sources that are detected
in each delay bin. We note that for SN\,1998bw-like and SN\,2009bb-like targets, the efficiency is 100\% regardless of the total time delay
(and thus, regardless of the adopted observing urgency strategy in the radio). 
In the case of SN\,2006aj-like targets, we completely miss those that are observed with total delays $\gtrsim 22$\,d,
when SN\,2006aj was undetected.

\begin{table}
\begin{center}
\caption{
Summary of our results for relativistic SNe. The efficiency quoted here is the average among all relativistic SNe listed in
Table \ref{tab:models}, for a high-urgency follow-up strategy. Targets that are not detectable have been excluded from this analysis.
See text for discussion.
}
\label{tab:relSN}
\begin{tabular}{ccc}
\hline\hline
$z$	& Efficiency	& days since 1$^{\rm st}$ obs.	\\
\hline
0.01	& 97\%	& 2, 8, 18, 30	\\
0.1	& 78\%	& 2, 6, 18, 34	\\
Mix	& 87\%	& 4, 10, 22, 30	\\
\hline
\end{tabular}
\end{center}
\end{table}

\subsubsection{Relevance of early-time radio observations for relativistic SNe}\label{sec:early}
In this Section we assess the importance of early-time radio observations for correctly and uniquely associating relativistic SNe via an
optimized radio follow-up campaign. To this end, we extrapolate the template light curves of SN\,2009bb and iPTF\,17cw using the early-time
behavior of the much better sampled radio light curves of SN\,1998bw and SN\,2006aj, respectively. We choose to use SN\,1998bw to
extrapolate the light curve of SN\,2009bb, and SN\,2006aj to extrapolate the light curve of iPTF\,17cw, because
they are the most similar to each other. In fact, both SN\,1998bw and SN\,2009bb were nearby explosions that resulted in a very bright radio
signal. On the other hand, both SN\,2006aj and iPTF\,17cw were very dim and faded away very rapidly.

We repeat the simulations and optimization procedure described in the previous Section including these early-time extrapolations.
Since all sources simulated based on the template of SN\,2009bb were both detected and correctly and uniquely associated without this
early-time extrapolation, we do not expect significant changes in the results for SN\,2009bb-like sources.
On the other hand, we do expect an improvement in the results for iPTF\,17cw-like SNe, specifically for the cases of $z=0.1$
and mixed redshifts (since all iPTF\,17cw-like sources were detected, and correctly and uniquely associated, at $z=0.01$). 

A summary of the results from this analysis, averaged over all the relativistic SN templates, is reported in Table~\ref{tab:relSN_early},
and confirm our expectations.
Specifically for iPTF\,17cw-like targets at $z=0.1$, 95\% are detected, and 64\% of those are correctly and uniquely associated.
For mixed redshifts, all of the iPTF\,17cw-like sources are detected, and 95\% of them are correctly
and uniquely associated.
These results highlight the need for early-time radio observations of newly-discovered relativistic SNe.
Not only a high-urgency strategy is indeed favored for triggering the first radio observation, but all of the follow-up campaign
should be conducted within the first 2-3 weeks since explosion.

\begin{table}
\begin{center}
\caption{Summary of our results for relativistic SNe when extrapolating the light curves of SN\,2009bb and iPTF\,17cw
to early times, as explained in Section~\ref{sec:early}.
The efficiency quoted here is the average among all relativistic SNe listed in Table \ref{tab:models}, for a high-urgency follow-up strategy.
Targets that are not detectable have been excluded from this analysis.
See text for discussion.
}
\label{tab:relSN_early}
\begin{tabular}{ccc}
\hline\hline
$z$	& Efficiency	& days since 1$^{\rm st}$ obs.	\\
\hline
0.01	& 97\%	& 4, 8, 14, 22	\\
0.1	& 83\%	& 2, 6, 14, 18	\\
Mix	& 91\%	& 2, 4, 8, 10	\\
\hline
\end{tabular}
\end{center}
\end{table}

\subsection{Discovering CSM-interacting SNe} \label{sec:CSMSN}
Our goal here is to optimize the observational strategy to detect CSM-interacting SNe in the radio. For this reason, we treat the CSM-interacting
SN templates listed in Table \ref{tab:models} as our targets, while relativistic SNe and off-axis GRBs are treated as contaminants.

The earliest detection for both our CSM-interacting templates happened around 30\,d after the explosion.
It is therefore expected that, in terms of urgency, we would find the highest number of positive identifications when adopting
a low-urgency radio follow-up, i.e., the largest interval between the optical discovery and the first radio observation
(see Section~\ref{sec:models} for how we extrapolate template fluxes at epochs preceding the earliest detection).
Moreover, as evident from Figure~\ref{fig:models}, the radio light curves of CSM-interacting SNe 
start diverging
from the light curves of
other contaminants around 40-70\,d after the explosion, so one can already expect that observations 
after these epochs
(which correspond to about 30-60\,d after the first radio observation) would be optimal.

We run our simulations with the same redshift intervals as for relativistic SNe in Section~\ref{sec:relSN}. 
Results are reported in Table~\ref{tab:CSMSN}.
In this case, we find that three epochs of radio follow-up observations are sufficient to correctly and uniquely identify all of the simulated
sources, and that using a low-urgency strategy suffices. Overall, the optimal radio follow-up strategy for CSM-interacting SNe requires
observations at later times than relativistic SNe. However, we also stress that the results reported here are subject to uncertainties related
to the limited number of radio-emitting, CSM-interacting BL-Ic SNe we know of so far. More discoveries of this type of explosions in the
future will enable us to better refine radio follow-up strategies.

\begin{table}
\begin{center}
\caption{Summary of our results for CSM-interacting SNe. The efficiency quoted here is the average among all CSM-interacting SNe
listed in Table \ref{tab:models}, for a low-urgency follow-up strategy. Targets that are not detectable have been excluded from this analysis.
See text for discussion.
}
\label{tab:CSMSN}
\begin{tabular}{ccc}
\hline
\hline
$z$	& Efficiency	& days since 1$^{\rm st}$ obs.	\\
\hline
0.01	& 100\%	& 6, 90	\\
0.1	& 100\%	& 22, 90	\\
Mix	& 100\%	& 6, 90	\\
\hline
\end{tabular}
\end{center}
\end{table}

\subsection{Discovering Off-Axis GRBs}
In this last Section our goal is to optimize the observational strategy to detect off-axis long GRBs. We therefore treat the off-axis GRB models
listed in Table \ref{tab:models} as our targets, while relativistic SN and the CSM-interacting SN templates are treated as contaminants.
As can be seen from Figure~\ref{fig:models}, off-axis GRB models span a variety of fluxes and timescales, and by construction
are much better sampled than our other SN templates at early and late times.  
Thus, we generally expect that a large fraction of the detectable sources will also be be uniquely and correctly identified.
However, we note that not all off-axis GRB models are detectable at all distances. In particular, the peak flux of the E48\_theta45 model
is below the radio detection threshold even at $z=0.01$. Moreover, the peak fluxes of the E49\_theta45 and the E48\_theta24 models
are barely above the radio detection threshold for a short time, so the radio follow-up efficiency is largely dominated by the delay
between explosion and optical discovery. 

Results of our simulations are reported in Table~\ref{tab:offaxis}.
We find that five epochs are necessary to maximize the amount of correct and unique associations, and that a high-urgency strategy is
preferable, especially for the low luminosity GRBs ($E_{\rm iso}<10^{51}\,$erg).

At $z=0.01$, overall 82\% of sources are detected.
In particular, 3\% of E48\_theta24, 23\% of E49\_theta24, and 39\% of E49\_theta45 are detected.
The fact, for sources with $E_{\rm iso}=10^{49}\,$erg, the case $\theta_v$=45\,deg yields to more detections than $\theta_v$=24\,deg
is explained by the faster evolution of E49\_theta24  which, in spite of having a peak flux brighter than E49\_theta45, becomes
quickly undetectable after less than 6 days since explosion.
For all other GRB model parameters, all sources are detected at $z=0.01$.
Overall, 99\% of the off-axis GRBs detected at $z=0.01$ are uniquely and correctly associated. Specifically, unique and correct
association efficiencies are 100\% for $E_{\rm iso}>10^{51}\,$erg, 81\% for E48\_theta24, 90\% for E49\_theta24, 99\% for E50\_theta24, 96\%
for E49\_theta45, and 100\% for E50\_theta45.

At $z=0.1$, overall 63\% of off-axis GRBs are detected. All explosions with $E_{\rm iso}>10^{51}\,$erg are always detectable;
25\% of E50\_theta24 are detectable; while sources with $E_{\rm iso}\le10^{50}\,$erg and $\theta_v$=45\,deg, and sources with
$E_{\rm iso}\le10^{49}\,$erg and $\theta_v$=24\,deg are never detectable at this redshift.
Overall, of the detectable sources at $z=0.1$, 96\% are uniquely and correctly associated.
Specifically, unique and correct association efficiencies are of 100\%
for all detectable sources with $\theta_v$=45\,deg,
4\% for E50\_theta24, 91\% for E51\_theta24,
and 100\% for explosions with $E_{\rm iso}>10^{52}\,$erg.

Finally, in the case of sources with redshift randomly distributed between 0.01 and 0.1 our results are, as expected, in between the two
previously described cases. Overall,  73\% of the off-axis GRBs are detected.
More specifically, 100\% of $E_{\rm iso}>10^{51}\,$erg, 0.4\% of E48\_theta24, 9\% of E49\_theta24, 58\% of E50\_theta24, 
9\% of E49\_theta45, and 70\% of E50\_theta45 are detected. Overall, 96\% of the detectable off-axis GRBs with mixed redshifts
are uniquely and correctly associated. In particular, unique and correct association efficiencies are as follows: 100\% for
$E_{\rm iso}>10^{52}\,$erg, 77\% for E48\_theta24, 73\% for E49\_theta24, 73\% for E50\_theta24, 98\% for E51\_theta24,
 68\% for E49\_theta45, 78\% for E50\_theta45, and 100\% for E51\_theta45.

\begin{table}
\begin{center}
\caption{Summary of our results for off-axis GRBs. The efficiency quoted here is the average among all off-axis GRB models
listed in Table \ref{tab:models}, for a high-urgency strategy. Targets that are not detectable have been excluded from this analysis.
See text for discussion.
}
\label{tab:offaxis}
\begin{tabular}{ccc}
\hline
\hline
$z$	& Efficiency	& days since 1$^{\rm st}$ obs.	\\
\hline
0.01	& 99\%	& 6, 22, 26, 82		\\
0.1	& 96\%	& 10, 14, 26, 82	\\
Mix	& 96\%	& 4, 10, 26, 82		\\
\hline
\end{tabular}
\end{center}
\end{table}

\section{Detectability in X-rays}\label{sec:xrays}

Hereafter we consider the benefits of X-ray follow-up observations 
of both relativistic SNe and CSM-interacting SNe, and the potential for X-ray detections.
Radio and X-ray observations both probe the fastest component of the SN ejecta. Combining radio and X-ray data one can independently constrain the density of the medium ($n_{\rm ISM}$) and the fraction of ejecta energy converted in
magnetic fields \citep[$\epsilon_{\rm B}$;][]{Chevalier2006}.

We estimate the X-ray flux of our targets at the time of the radio peak, assuming the X-rays are produced via synchrotron emission
with radio-to-X-ray spectrum defined as follows:
\begin{equation}
F_{X}=F_{\rm radio}\times\left(\frac{\nu_{ X}}{\nu_{\rm radio}}\right)^{-\beta},
\label{eq:1}
\end{equation}
where $F_{ X}$ and $F_{\rm radio}$ are the fluxes in the X-ray and
radio bands respectively, $\nu_{\rm X}$ and $\nu_{\rm radio}$ are the frequencies of the X-ray and radio observations respectively,
and $\beta\approx0.7-1$ is the spectral index.

We test the detectability of the X-ray emission from our sources with both {\emph {Swift}} and {\emph {Chandra}}.
For what concerns X-ray observations with {\emph {Swift}}, in a $\sim$10\,ks-long observation one can reach a 3$\sigma$ sensitivity of
$\sim$2.5$\times$10$^{-14}$\,erg\,cm$^{-2}$\,s$^{-1}$ \citep[unabsorbed flux;][]{Gehrels2004}.
With $\beta$=1, all sources would be too dim to be detectable.
On the other hand, with $\beta$=0.7, all sources would be detectable at $z=0.01$, but none would be detectable at $z=0.1$.
With a 20\,ks-long observation with {\emph {Chandra}} one could reach a 3$\sigma$ sensitivity of $\sim$3$\times$10$^{-15}$
erg\,cm$^{-2}$\,s$^{-1}$  \citep[unabsorbed flux;][]{Burrows2005}.
In this case, with $\beta$=1, only SN\,1998bw, SN\,2009bb, PTF\,2011qcj, and SN\,2007bg would be detectable at $z=0.01$,
while none would be at $z=0.1$.
With $\beta$=0.7, all sources would be detectable even at $z=0.1$, although SN\,2006aj and PTF\,2017cw would be very close to the
detection threshold.

We also calculate for how long the X-ray emission would be detectable by \textit{Chandra}.
Our results are reported in Table~\ref{tab:Xray_time}.
These results assume that, during the whole time, the X-ray emission is produced via synchrotron radiation with
a constant radio-to-X-ray spectral index as in Equation~\ref{eq:1}.
We note that for all sources, with $\beta$=0.7 and $z$=0.01, the X-ray emission would be visible at least as long as we have radio
observations of the sources.

We finally calculate the distance limit (i.e. the distance at which the flux would be equal to the sensitivity limit of \textit{Chandra})
for each source. Our results are reported in Table~\ref{tab:Xray_dist}.
We highlight that the distance limits derived for SN\,2006aj-like and iPTF17cw-like SNe in the case $\beta$=1 are closer than the
actual distance to these sources, despite both of them were detected in X-rays \citep[][]{Campana2006,Corsi2017}.
This is explained by the fact that evidence for a flattening of the radio-to-X-ray spectral index, possibly related to cosmic-ray dominated shocks,
has been observed in these events \citep{Ellison2000, Chevalier2006,Vink2017}. 
The distance limit we obtain for SN\,2007bg is also closer than the source's actual distance (152\,Mpc),
in agreement with the fact that no X-ray detection was reported \citep[$F_X<2.6\times10^{-15}$\,erg\,cm$^{-2}$\,s$^{-1}$;][]{Salas2013}.

\begin{table}
\begin{center}
\caption{
Maximum time after explosion at which X-ray emission would be detectable in 
a 20\,ks-long observation with {\emph {Chandra}}.
We assume X-rays are produced via synchrotron emission with constant radio-to-X-ray spectral index
in the range $\beta\approx 0.7-1$ over these timescales.
The first entry corresponds to the case $\beta=0.7$, while the second to $\beta=1$.
A forward slash indicates that a source is never detectable. A greater than ($>$) symbol indicates that the source is detectable at least until the time of the latest radio observation we considered. 
See text for discussion.}
\label{tab:Xray_time}
\begin{tabular}{ccc}
\hline
\hline
Source	&  $\Delta t_{{\rm max}, z=0.01}$ & $\Delta t_{{\rm max}, z=0.1}$	\\
 & (days) & (days) \\
\hline
SN\,1998bw	& $>250-\sim70$	& $\sim90-/$		\\
SN\,2009bb	& $>150-\sim70$	& $\sim100-/$	\\
SN\,2006aj	& $> 22-/$			& $\sim5-/$		\\
iPTF\,17cw	& $>31-/$			& $\sim16-/$		\\
PTF\,11qcj	& $>600->600$		& $>600-/$		\\
SN\,2007bg	& $>860->860$		& $>860-/$		\\
\hline
\end{tabular}
\end{center}
\end{table}

\begin{table}
\begin{center}
\caption{Distance at which the X-ray flux of both relativistic and CSM-interacting SNe 
would be equal to the $3\sigma$ sensitivity limit of
a 20\,ks-long observation with {\emph{Chandra}},
assuming the radio-to-X-ray emission is synchrotron radiation with no spectral breaks
and $\beta\approx 0.7-1$.
The first entry corresponds to the case $\beta=0.7$, while the second to $\beta=1$. See text for discussion.}
\label{tab:Xray_dist}
\begin{tabular}{cc}
\hline
\hline
Source	& Horizon 	\\
& (Mpc)\\
\hline
SN\,1998bw	& 1634-104	\\
SN\,2009bb	& 950-60	\\
SN\,2006aj	& 525-33	\\
iPTF\,17cw	& 567-36	\\
\hline
PTF11qcj		& 1853-117	\\
SN\,2007bg	& 1467-93	\\
\hline
\end{tabular}
\end{center}
\end{table}

\section{Summary and conclusion}\label{sec:conclusions}
We have presented an analysis aimed at identifying an optimal strategy for detecting and characterizing various types of
radio-emitting stripped-envelope core-collapse SNe with the VLA.

Our results show how early-time ($<$7 days after the explosion) radio observations are key to identifying relativistic,
engine-driven SNe, whose radio emission peaks early and fades away quickly. This is clearly demonstrated by our results for
SN\,2009bb-like and iPTF\,17cw-like explosions.
Radio emission from CSM-interacting SNe is typically longer-lived, and can successfully be identified via later times observations,
around 40-90 days after the explosion, although this conclusion is affected by uncertainties related to the limited number of
radio-emitting, CSM-interacting BL-Ic SNe we know of so  far.
For radio afterglows of off-axis long GRBs, early-time observations are required in order to maximize the probability of correctly interpreting their
origin and physical properties.

Finally, we discussed the detectability of relativistic and CSM-interacting SNe in X-rays. We found that, if their X-ray emission
is due to synchrotron radiation,
most of them are only detectable when they are relatively nearby ($<$100\,Mpc)
for spectral indices greater than unity, while they may be detected up to about 1\,Gpc for spectral indices of about 0.7.

In the near future, LSST will discover about $10^4$ BL-Ic SNe per year \citep[][]{LSST2009,Shivvers2017},
providing a fantastic resource to investigate the fraction of these events linked to long GRBs.
Utilizing an optimized strategy to follow-up BL-Ic SNe in the radio will be crucial to investigate as many events as 
possible, and put tighter constraints on the open question of the nature of their progenitors.
At the time LSST will be starting operations (mid-late 2020s), a next generation Very Large Array (ngVLA) will likely be starting
operations as well \citep{Murphy2018}.
ngVLA is a proposed next generation radio interferometer with $\approx 10\times$ the sensitivity of the current VLA, which will enable
discovery of sources $\approx 3\times$ as far, therefore enlarging the number of possible detections by about a factor of 30, and
dramatically expanding the capabilities to discover new radio-loud SNe of the rarest types.

\acknowledgments
A.C. and D.C. acknowledge support from the National Science Foundation CAREER Award \#1455090 (PI: A. Corsi).
A.C. and D.C. also acknowledges partial support from Chandra Award No. GO8-19055A.
iPTF\,17cw was discovered as part of the GROWTH project, which is funded by the National Science Foundation under Grant \# 1545949.
The National Radio Astronomy Observatory is a facility of the National Science Foundation operated under cooperative agreement
by Associated Universities, Inc.

\bibliographystyle{apj}
\bibliography{./bibliography.bib}

\begin{thebibliography}{}
\expandafter\ifx\csname natexlab\endcsname\relax\def\natexlab#1{#1}\fi

\bibitem[{{Adelman-McCarthy} {et~al.}(2006){Adelman-McCarthy}, {Ag{\"u}eros},
  {Allam}, {Anderson}, {Anderson}, {Annis}, {Bahcall}, {Baldry}, {Barentine},
  {Berlind}, {Bernardi}, {Blanton}, {Boroski}, {Brewington}, {Brinchmann},
  {Brinkmann}, {Brunner}, {Budav{\'a}ri}, {Carey}, {Carr}, {Castander},
  {Connolly}, {Csabai}, {Czarapata}, {Dalcanton}, {Doi}, {Dong}, {Eisenstein},
  {Evans}, {Fan}, {Finkbeiner}, {Friedman}, {Frieman}, {Fukugita}, {Gillespie},
  {Glazebrook}, {Gray}, {Grebel}, {Gunn}, {Gurbani}, {de Haas}, {Hall},
  {Harris}, {Harvanek}, {Hawley}, {Hayes}, {Hendry}, {Hennessy}, {Hindsley},
  {Hirata}, {Hogan}, {Hogg}, {Holmgren}, {Holtzman}, {Ichikawa}, {Ivezi{\'c}},
  {Jester}, {Johnston}, {Jorgensen}, {Juri{\'c}}, {Kent}, {Kleinman}, {Knapp},
  {Kniazev}, {Kron}, {Krzesinski}, {Kuropatkin}, {Lamb}, {Lampeitl}, {Lee},
  {Leger}, {Lin}, {Long}, {Loveday}, {Lupton}, {Margon},
  {Mart{\'{\i}}nez-Delgado}, {Mandelbaum}, {Matsubara}, {McGehee}, {McKay},
  {Meiksin}, {Munn}, {Nakajima}, {Nash}, {Neilsen}, {Newberg}, {Newman},
  {Nichol}, {Nicinski}, {Nieto-Santisteban}, {Nitta}, {O'Mullane}, {Okamura},
  {Owen}, {Padmanabhan}, {Pauls}, {Peoples}, {Pier}, {Pope}, {Pourbaix},
  {Quinn}, {Richards}, {Richmond}, {Rockosi}, {Schlegel}, {Schneider},
  {Schroeder}, {Scranton}, {Seljak}, {Sheldon}, {Shimasaku}, {Smith}, {Smol{\v
  c}i{\'c}}, {Snedden}, {Stoughton}, {Strauss}, {SubbaRao}, {Szalay},
  {Szapudi}, {Szkody}, {Tegmark}, {Thakar}, {Tucker}, {Uomoto}, {Vanden Berk},
  {Vandenberg}, {Vogeley}, {Voges}, {Vogt}, {Walkowicz}, {Weinberg}, {West},
  {White}, {Xu}, {Yanny}, {Yocum}, {York}, {Zehavi}, {Zibetti}, \&
  {Zucker}}]{SDSS_IV2006}
{Adelman-McCarthy}, J.~K., {Ag{\"u}eros}, M.~A., {Allam}, S.~S., {et~al.} 2006,
  \apjs, 162, 38

\bibitem[{{Amati}(2006)}]{Amati2006}
{Amati}, L. 2006, \mnras, 372, 233

\bibitem[{{Amati} {et~al.}(2002){Amati}, {Frontera}, {Tavani}, {in't Zand},
  {Antonelli}, {Costa}, {Feroci}, {Guidorzi}, {Heise}, {Masetti}, {Montanari},
  {Nicastro}, {Palazzi}, {Pian}, {Piro}, \& {Soffitta}}]{Amati2002}
{Amati}, L., {Frontera}, F., {Tavani}, M., {et~al.} 2002, \aap, 390, 81

\bibitem[{{Bellm}(2016)}]{Bellm2016}
{Bellm}, E.~C. 2016, \pasp, 128, 084501

\bibitem[{{Beniamini} \& {van der Horst}(2017)}]{Beniamini2017}
{Beniamini}, P., \& {van der Horst}, A.~J. 2017, \mnras, 472, 3161

\bibitem[{{Berger} {et~al.}(2003){Berger}, {Kulkarni}, {Frail}, \&
  {Soderberg}}]{Berger2003}
{Berger}, E., {Kulkarni}, S.~R., {Frail}, D.~A., \& {Soderberg}, A.~M. 2003,
  \apj, 599, 408

\bibitem[{{Bromberg} {et~al.}(2011){Bromberg}, {Nakar}, \&
  {Piran}}]{Bromberg2011}
{Bromberg}, O., {Nakar}, E., \& {Piran}, T. 2011, \apjl, 739, L55

\bibitem[{Burrows {et~al.}(2005)Burrows, Hill, Nousek, Kennea, Wells, Osborne,
  Abbey, Beardmore, Mukerjee, Short, Chincarini, Campana, Citterio, Moretti, \&
  et~al. Pagani}]{Burrows2005}
Burrows, D.~N., Hill, J.~E., Nousek, J.~A., {et~al.} 2005, \ssr, 120, 165

\bibitem[{{Campana} {et~al.}(2006){Campana}, {Mangano}, {Blustin}, {Brown},
  {Burrows}, {Chincarini}, {Cummings}, {Cusumano}, {Della Valle}, {Malesani},
  {M{\'e}sz{\'a}ros}, {Nousek}, {Page}, {Sakamoto}, {Waxman}, {Zhang}, {Dai},
  {Gehrels}, {Immler}, {Marshall}, {Mason}, {Moretti}, {O'Brien}, {Osborne},
  {Page}, {Romano}, {Roming}, {Tagliaferri}, {Cominsky}, {Giommi}, {Godet},
  {Kennea}, {Krimm}, {Angelini}, {Barthelmy}, {Boyd}, {Palmer}, {Wells}, \&
  {White}}]{Campana2006}
{Campana}, S., {Mangano}, V., {Blustin}, A.~J., {et~al.} 2006, \nat, 442, 1008

\bibitem[{{Carbone} \& {Corsi}(2018)}]{Carbone2018}
{Carbone}, D., \& {Corsi}, A. 2018, \apj, 867, 135

\bibitem[{{Chandra} \& {Frail}(2012)}]{Chandra2012}
{Chandra}, P., \& {Frail}, D.~A. 2012, \apj, 746, 156

\bibitem[{{Chevalier} \& {Fransson}(2006)}]{Chevalier2006}
{Chevalier}, R.~A., \& {Fransson}, C. 2006, \apj, 651, 381

\bibitem[{{Corsi} {et~al.}(2011){Corsi}, {Ofek}, {Frail}, {Poznanski},
  {Arcavi}, {Gal-Yam}, {Kulkarni}, {Hurley}, {Mazzali}, {Howell}, {Kasliwal},
  {Green}, {Murray}, {Sullivan}, {Xu}, {Ben-ami}, {Bloom}, {Cenko}, {Law},
  {Nugent}, {Quimby}, {Pal'shin}, {Cummings}, {Connaughton}, {Yamaoka}, {Rau},
  {Boynton}, {Mitrofanov}, \& {Goldsten}}]{Corsi2011}
{Corsi}, A., {Ofek}, E.~O., {Frail}, D.~A., {et~al.} 2011, \apj, 741, 76

\bibitem[{{Corsi} {et~al.}(2014){Corsi}, {Ofek}, {Gal-Yam}, {Frail},
  {Kulkarni}, {Fox}, {Kasliwal}, {Sullivan}, {Horesh}, {Carpenter}, {Maguire},
  {Arcavi}, {Cenko}, {Cao}, {Mooley}, {Pan}, {Sesar}, {Sternberg}, {Xu},
  {Bersier}, {James}, {Bloom}, \& {Nugent}}]{Corsi2014}
{Corsi}, A., {Ofek}, E.~O., {Gal-Yam}, A., {et~al.} 2014, \apj, 782, 42

\bibitem[{{Corsi} {et~al.}(2016){Corsi}, {Gal-Yam}, {Kulkarni}, {Frail},
  {Mazzali}, {Cenko}, {Kasliwal}, {Cao}, {Horesh}, {Palliyaguru}, {Perley},
  {Laher}, {Taddia}, {Leloudas}, {Maguire}, {Nugent}, {Sollerman}, \&
  {Sullivan}}]{Corsi2016}
{Corsi}, A., {Gal-Yam}, A., {Kulkarni}, S.~R., {et~al.} 2016, \apj, 830, 42

\bibitem[{{Corsi} {et~al.}(2017){Corsi}, {Cenko}, {Kasliwal}, {Quimby},
  {Kulkarni}, {Frail}, {Goldstein}, {Blagorodnova}, {Connaughton}, {Perley},
  {Singer}, {Copperwheat}, {Fremling}, {Kupfer}, {Piascik}, {Steele}, {Taddia},
  {Vedantham}, {Kutyrev}, {Palliyaguru}, {Roberts}, {Sollerman}, {Troja}, \&
  {Veilleux}}]{Corsi2017}
{Corsi}, A., {Cenko}, S.~B., {Kasliwal}, M.~M., {et~al.} 2017, \apj, 847, 54

\bibitem[{{Eichler} \& {Levinson}(1999)}]{Eichler1999}
{Eichler}, D., \& {Levinson}, A. 1999, \apjl, 521, L117

\bibitem[{{Ellison} {et~al.}(2000){Ellison}, {Berezhko}, \&
  {Baring}}]{Ellison2000}
{Ellison}, D.~C., {Berezhko}, E.~G., \& {Baring}, M.~G. 2000, \apj, 540, 292

\bibitem[{{Filippenko}(1997)}]{Filippenko1997}
{Filippenko}, A.~V. 1997, \araa, 35, 309

\bibitem[{{Foley} {et~al.}(2006){Foley}, {Watson}, {Gorosabel}, {Fynbo},
  {Sollerman}, {McGlynn}, {McBreen}, \& {Hjorth}}]{Foley2006}
{Foley}, S., {Watson}, D., {Gorosabel}, J., {et~al.} 2006, \aap, 447, 891

\bibitem[{{Frail} {et~al.}(2001){Frail}, {Kulkarni}, {Sari}, {Djorgovski},
  {Bloom}, {Galama}, {Reichart}, {Berger}, {Harrison}, {Price}, {Yost},
  {Diercks}, {Goodrich}, \& {Chaffee}}]{Frail2001}
{Frail}, D.~A., {Kulkarni}, S.~R., {Sari}, R., {et~al.} 2001, \apjl, 562, L55

\bibitem[{{Gal-Yam}(2017)}]{GalYam2017}
{Gal-Yam}, A. 2017, {Observational and Physical Classification of Supernovae},
  195

\bibitem[{{Galama} {et~al.}(1998){Galama}, {Vreeswijk}, {van Paradijs},
  {Kouveliotou}, {Augusteijn}, {B{\"o}hnhardt}, {Brewer}, {Doublier},
  {Gonzalez}, {Leibundgut}, {Lidman}, {Hainaut}, {Patat}, {Heise}, {in't Zand},
  {Hurley}, {Groot}, {Strom}, {Mazzali}, {Iwamoto}, {Nomoto}, {Umeda},
  {Nakamura}, {Young}, {Suzuki}, {Shigeyama}, {Koshut}, {Kippen}, {Robinson},
  {de Wildt}, {Wijers}, {Tanvir}, {Greiner}, {Pian}, {Palazzi}, {Frontera},
  {Masetti}, {Nicastro}, {Feroci}, {Costa}, {Piro}, {Peterson}, {Tinney},
  {Boyle}, {Cannon}, {Stathakis}, {Sadler}, {Begam}, \& {Ianna}}]{Galama1998}
{Galama}, T.~J., {Vreeswijk}, P.~M., {van Paradijs}, J., {et~al.} 1998, \nat,
  395, 670

\bibitem[{Gehrels {et~al.}(2004)Gehrels, Chincarini, Giommi, Mason, Nousek,
  Wells, White, Barthelmy, Burrows, Cominsky, Hurley, Marshall,
  M\'{e}sz\'{a}ros, Roming, Angelini, \& et~al. Barbier}]{Gehrels2004}
Gehrels, N., Chincarini, G., Giommi, P., {et~al.} 2004, \apj, 611, 1005

\bibitem[{{Ghirlanda} {et~al.}(2005){Ghirlanda}, {Ghisellini}, \&
  {Firmani}}]{Ghirlanda2005}
{Ghirlanda}, G., {Ghisellini}, G., \& {Firmani}, C. 2005, \mnras, 361, L10

\bibitem[{{Ghirlanda} {et~al.}(2004){Ghirlanda}, {Ghisellini}, \&
  {Lazzati}}]{Ghirlanda2004}
{Ghirlanda}, G., {Ghisellini}, G., \& {Lazzati}, D. 2004, \apj, 616, 331

\bibitem[{{Goldstein} {et~al.}(2016){Goldstein}, {Connaughton}, {Briggs}, \&
  {Burns}}]{Goldstein2016}
{Goldstein}, A., {Connaughton}, V., {Briggs}, M.~S., \& {Burns}, E. 2016, \apj,
  818, 18

\bibitem[{{Graham} {et~al.}(2019){Graham}, {Kulkarni}, {Bellm}, {Adams},
  {Barbarino}, {Blagorodnova}, {Bodewits}, {Bolin}, {Brady}, {Cenko}, {Chang},
  {Coughlin}, {De}, {Eadie}, {Farnham}, {Feindt}, {Franckowiak}, {Fremling},
  {Gezari}, {Ghosh}, {Goldstein}, {Golkhou}, {Goobar}, {Ho}, {Huppenkothen},
  {Ivezi{\'c}}, {Jones}, {Juric}, {Kaplan}, {Kasliwal}, {Kelley}, {Kupfer},
  {Lee}, {Lin}, {Lunnan}, {Mahabal}, {Miller}, {Ngeow}, {Nugent}, {Ofek},
  {Prince}, {Rauch}, {van Roestel}, {Schulze}, {Singer}, {Sollerman}, {Taddia},
  {Yan}, {Ye}, {Yu}, {Barlow}, {Bauer}, {Beck}, {Belicki}, {Biswas}, {Brinnel},
  {Brooke}, {Bue}, {Bulla}, {Burruss}, {Connolly}, {Cromer}, {Cunningham},
  {Dekany}, {Delacroix}, {Desai}, {Duev}, {Feeney}, {Flynn}, {Frederick},
  {Gal-Yam}, {Giomi}, {Groom}, {Hacopians}, {Hale}, {Helou}, {Henning},
  {Hover}, {Hillenbrand}, {Howell}, {Hung}, {Imel}, {Ip}, {Jackson}, {Kaspi},
  {Kaye}, {Kowalski}, {Kramer}, {Kuhn}, {Landry}, {Laher}, {Mao}, {Masci},
  {Monkewitz}, {Murphy}, {Nordin}, {Patterson}, {Penprase}, {Porter},
  {Rebbapragada}, {Reiley}, {Riddle}, {Rigault}, {Rodriguez}, {Rusholme}, {van
  Santen}, {Shupe}, {Smith}, {Soumagnac}, {Stein}, {Surace}, {Szkody}, {Terek},
  {Van Sistine}, {van Velzen}, {Vestrand}, {Walters}, {Ward}, {Zhang}, \&
  {Zolkower}}]{Graham2019}
{Graham}, M.~J., {Kulkarni}, S.~R., {Bellm}, E.~C., {et~al.} 2019, \pasp, 131,
  078001

\bibitem[{{Granot} \& {van der Horst}(2014)}]{Granot2014}
{Granot}, J., \& {van der Horst}, A.~J. 2014, \pasa, 31, e008

\bibitem[{{Ho} {et~al.}(2019){Ho}, {Goldstein}, {Schulze}, {Khatami}, {Perley},
  {Ergon}, {Gal-Yam}, {Corsi}, {Andreoni}, {Barbarino}, {Bellm},
  {Blagorodnova}, {Bright}, {Burns}, {Cenko}, {Cunningham}, {De}, {Dekany},
  {Dugas}, {Fender}, {Fransson}, {Fremling}, {Goldstein}, {Graham}, {Hale},
  {Horesh}, {Hung}, {Kasliwal}, {Kuin}, {Kulkarni}, {Kupfer}, {Lunnan},
  {Masci}, {Ngeow}, {Nugent}, {Ofek}, {Patterson}, {Petitpas}, {Rusholme},
  {Sai}, {Sfaradi}, {Shupe}, {Sollerman}, {Soumagnac}, {Tachibana}, {Taddia},
  {Walters}, {Wang}, {Yao}, \& {Zhang}}]{Ho2019}
{Ho}, A.~Y.~Q., {Goldstein}, D.~A., {Schulze}, S., {et~al.} 2019, arXiv
  e-prints, arXiv:1904.11009

\bibitem[{{Kellermann}(1964)}]{Kellermann1964}
{Kellermann}, K.~I. 1964, \apj, 140, 969

\bibitem[{{Kulkarni} {et~al.}(1998){Kulkarni}, {Frail}, {Wieringa}, {Ekers},
  {Sadler}, {Wark}, {Higdon}, {Phinney}, \& {Bloom}}]{Kulkarni1998}
{Kulkarni}, S.~R., {Frail}, D.~A., {Wieringa}, M.~H., {et~al.} 1998, \nat, 395,
  663

\bibitem[{{Law} {et~al.}(2009){Law}, {Kulkarni}, {Dekany}, {Ofek}, {Quimby},
  {Nugent}, {Surace}, {Grillmair}, {Bloom}, {Kasliwal}, {Bildsten}, {Brown},
  {Cenko}, {Ciardi}, {Croner}, {Djorgovski}, {van Eyken}, {Filippenko}, {Fox},
  {Gal-Yam}, {Hale}, {Hamam}, {Helou}, {Henning}, {Howell}, {Jacobsen},
  {Laher}, {Mattingly}, {McKenna}, {Pickles}, {Poznanski}, {Rahmer}, {Rau},
  {Rosing}, {Shara}, {Smith}, {Starr}, {Sullivan}, {Velur}, {Walters}, \&
  {Zolkower}}]{Law2009}
{Law}, N.~M., {Kulkarni}, S.~R., {Dekany}, R.~G., {et~al.} 2009, \pasp, 121,
  1395

\bibitem[{{LSST Science Collaboration} {et~al.}(2009){LSST Science
  Collaboration}, {Abell}, {Allison}, {Anderson}, {Andrew}, {Angel}, {Armus},
  {Arnett}, {Asztalos}, \& {Axelrod}}]{LSST2009}
{LSST Science Collaboration}, {Abell}, P.~A., {Allison}, J., {et~al.} 2009,
  arXiv e-prints, arXiv:0912.0201

\bibitem[{{Margutti} {et~al.}(2019){Margutti}, {Metzger}, {Chornock}, {Vurm},
  {Roth}, {Grefenstette}, {Savchenko}, {Cartier}, {Steiner}, {Terreran},
  {Margalit}, {Migliori}, {Milisavljevic}, {Alexander}, {Bietenholz},
  {Blanchard}, {Bozzo}, {Brethauer}, {Chilingarian}, {Coppejans}, {Ducci},
  {Ferrigno}, {Fong}, {G{\"o}tz}, {Guidorzi}, {Hajela}, {Hurley}, {Kuulkers},
  {Laurent}, {Mereghetti}, {Nicholl}, {Patnaude}, {Ubertini}, {Banovetz},
  {Bartel}, {Berger}, {Coughlin}, {Eftekhari}, {Frederiks}, {Kozlova},
  {Laskar}, {Svinkin}, {Drout}, {MacFadyen}, \& {Paterson}}]{Margutti2019}
{Margutti}, R., {Metzger}, B.~D., {Chornock}, R., {et~al.} 2019, \apj, 872, 18

\bibitem[{{Mazzali} {et~al.}(2014){Mazzali}, {McFadyen}, {Woosley}, {Pian}, \&
  {Tanaka}}]{Mazzali2014}
{Mazzali}, P.~A., {McFadyen}, A.~I., {Woosley}, S.~E., {Pian}, E., \& {Tanaka},
  M. 2014, \mnras, 443, 67

\bibitem[{{McMullin} {et~al.}(2007){McMullin}, {Waters}, {Schiebel}, {Young},
  \& {Golap}}]{McMullin2007}
{McMullin}, J.~P., {Waters}, B., {Schiebel}, D., {Young}, W., \& {Golap}, K.
  2007, in Astronomical Society of the Pacific Conference Series, Vol. 376,
  Astronomical Data Analysis Software and Systems XVI, ed. R.~A. {Shaw},
  F.~{Hill}, \& D.~J. {Bell}, 127

\bibitem[{{Murphy} {et~al.}(2018){Murphy}, {Bolatto}, {Chatterjee}, {Casey},
  {Chomiuk}, {Dale}, {de Pater}, {Dickinson}, {Francesco}, {Hallinan},
  {Isella}, {Kohno}, {Kulkarni}, {Lang}, {Lazio}, {Leroy}, {Loinard},
  {Maccarone}, {Matthews}, {Osten}, {Reid}, {Riechers}, {Sakai}, {Walter}, \&
  {Wilner}}]{Murphy2018}
{Murphy}, E.~J., {Bolatto}, A., {Chatterjee}, S., {et~al.} 2018, in
  Astronomical Society of the Pacific Conference Series, Vol. 517, Science with
  a Next Generation Very Large Array, ed. E.~{Murphy}, 3

\bibitem[{{Nava} {et~al.}(2012){Nava}, {Salvaterra}, {Ghirlanda}, {Ghisellini},
  {Campana}, {Covino}, {Cusumano}, {D'Avanzo}, {D'Elia}, \&
  {Fugazza}}]{Nava2012}
{Nava}, L., {Salvaterra}, R., {Ghirlanda}, G., {et~al.} 2012, \mnras, 421, 1256

\bibitem[{{Palliyaguru} {et~al.}(2019){Palliyaguru}, {Corsi}, {Frail},
  {Vink{\'o}}, {Wheeler}, {Gal-Yam}, {Cenko}, {Kulkarni}, \&
  {Kasliwal}}]{Palliyaguru2019}
{Palliyaguru}, N.~T., {Corsi}, A., {Frail}, D.~A., {et~al.} 2019, \apj, 872,
  201

\bibitem[{{Perley} {et~al.}(2014){Perley}, {Cenko}, {Corsi}, {Tanvir}, {Levan},
  {Kann}, {Sonbas}, {Wiersema}, {Zheng}, {Zhao}, {Bai}, {Bremer},
  {Castro-Tirado}, {Chang}, {Clubb}, {Frail}, {Fruchter}, {G{\"o}{\u g}{\"u}{\c
  s}}, {Greiner}, {G{\"u}ver}, {Horesh}, {Filippenko}, {Klose}, {Mao},
  {Morgan}, {Pozanenko}, {Schmidl}, {Stecklum}, {Tanga}, {Volnova}, {Volvach},
  {Wang}, {Winters}, \& {Xin}}]{Perley2014}
{Perley}, D.~A., {Cenko}, S.~B., {Corsi}, A., {et~al.} 2014, \apj, 781, 37

\bibitem[{{Perna} \& {Loeb}(1998)}]{Perna1998}
{Perna}, R., \& {Loeb}, A. 1998, \apjl, 509, L85

\bibitem[{{Piran}(2004)}]{Piran2004}
{Piran}, T. 2004, Reviews of Modern Physics, 76, 1143

\bibitem[{{Prentice} {et~al.}(2018){Prentice}, {Maguire}, {Smartt}, {Magee},
  {Schady}, {Sim}, {Chen}, {Clark}, {Colin}, {Fulton}, {McBrien}, {O'Neill},
  {Smith}, {Ashall}, {Chambers}, {Denneau}, {Flewelling}, {Heinze}, {Holoien},
  {Huber}, {Kochanek}, {Mazzali}, {Prieto}, {Rest}, {Shappee}, {Stalder},
  {Stanek}, {Stritzinger}, {Thompson}, \& {Tonry}}]{Prentice2018}
{Prentice}, S.~J., {Maguire}, K., {Smartt}, S.~J., {et~al.} 2018, \apjl, 865,
  L3

\bibitem[{{Rhoads}(1997)}]{Rhoads1997}
{Rhoads}, J.~E. 1997, \apjl, 487, L1

\bibitem[{{Salas} {et~al.}(2013){Salas}, {Bauer}, {Stockdale}, \&
  {Prieto}}]{Salas2013}
{Salas}, P., {Bauer}, F.~E., {Stockdale}, C., \& {Prieto}, J.~L. 2013, \mnras,
  428, 1207

\bibitem[{{Shivvers} {et~al.}(2017){Shivvers}, {Modjaz}, {Zheng}, {Liu},
  {Filippenko}, {Silverman}, {Matheson}, {Pastorello}, {Graur}, \&
  {Foley}}]{Shivvers2017}
{Shivvers}, I., {Modjaz}, M., {Zheng}, W., {et~al.} 2017, \pasp, 129, 054201

\bibitem[{{Smith} {et~al.}(2014){Smith}, {Dekany}, {Bebek}, {Bellm}, {Bui},
  {Cromer}, {Gardner}, {Hoff}, {Kaye}, {Kulkarni}, {Lambert}, {Levi}, \&
  {Reiley}}]{Smith2014}
{Smith}, R.~M., {Dekany}, R.~G., {Bebek}, C., {et~al.} 2014, in \procspie, Vol.
  9147, Ground-based and Airborne Instrumentation for Astronomy V, 914779

\bibitem[{{Soderberg} {et~al.}(2006{\natexlab{a}}){Soderberg}, {Chevalier},
  {Kulkarni}, \& {Frail}}]{Soderberg2006b}
{Soderberg}, A.~M., {Chevalier}, R.~A., {Kulkarni}, S.~R., \& {Frail}, D.~A.
  2006{\natexlab{a}}, \apj, 651, 1005

\bibitem[{{Soderberg} {et~al.}(2006{\natexlab{b}}){Soderberg}, {Nakar},
  {Berger}, \& {Kulkarni}}]{Soderberg2006c}
{Soderberg}, A.~M., {Nakar}, E., {Berger}, E., \& {Kulkarni}, S.~R.
  2006{\natexlab{b}}, \apj, 638, 930

\bibitem[{{Soderberg} {et~al.}(2006{\natexlab{c}}){Soderberg}, {Kulkarni},
  {Nakar}, {Berger}, {Cameron}, {Fox}, {Frail}, {Gal-Yam}, {Sari}, {Cenko},
  {Kasliwal}, {Chevalier}, {Piran}, {Price}, {Schmidt}, {Pooley}, {Moon},
  {Penprase}, {Ofek}, {Rau}, {Gehrels}, {Nousek}, {Burrows}, {Persson}, \&
  {McCarthy}}]{Soderberg2006}
{Soderberg}, A.~M., {Kulkarni}, S.~R., {Nakar}, E., {et~al.}
  2006{\natexlab{c}}, \nat, 442, 1014

\bibitem[{{Soderberg} {et~al.}(2010){Soderberg}, {Chakraborti}, {Pignata},
  {Chevalier}, {Chandra}, {Ray}, {Wieringa}, {Copete}, {Chaplin},
  {Connaughton}, {Barthelmy}, {Bietenholz}, {Chugai}, {Stritzinger}, {Hamuy},
  {Fransson}, {Fox}, {Levesque}, {Grindlay}, {Challis}, {Foley}, {Kirshner},
  {Milne}, \& {Torres}}]{Soderberg2010}
{Soderberg}, A.~M., {Chakraborti}, S., {Pignata}, G., {et~al.} 2010, \nat, 463,
  513

\bibitem[{{Sollerman} {et~al.}(2006){Sollerman}, {Jaunsen}, {Fynbo}, {Hjorth},
  {Jakobsson}, {Stritzinger}, {F{\'e}ron}, {Laursen}, {Ovaldsen}, {Selj},
  {Th{\"o}ne}, {Xu}, {Davis}, {Gorosabel}, {Watson}, {Duro}, {Ilyin}, {Jensen},
  {Lysfjord}, {Marquart}, {Nielsen}, {N{\"a}r{\"a}nen}, {Schwarz}, {Walch},
  {Wold}, \& {{\"O}stlin}}]{Sollerman2006}
{Sollerman}, J., {Jaunsen}, A.~O., {Fynbo}, J.~P.~U., {et~al.} 2006, \aap, 454,
  503

\bibitem[{{Strauss} {et~al.}(1992){Strauss}, {Huchra}, {Davis}, {Yahil},
  {Fisher}, \& {Tonry}}]{Strauss1992}
{Strauss}, M.~A., {Huchra}, J.~P., {Davis}, M., {et~al.} 1992, \apjs, 83, 29

\bibitem[{{van Eerten} {et~al.}(2012){van Eerten}, {van der Horst}, \&
  {MacFadyen}}]{vanEerten2012}
{van Eerten}, H., {van der Horst}, A., \& {MacFadyen}, A. 2012, \apj, 749, 44

\bibitem[{{Vink}(2017)}]{Vink2017}
{Vink}, J. 2017, {X-Ray Emission Properties of Supernova Remnants}, ed. A.~W.
  {Alsabti} \& P.~{Murdin}, 2063

\bibitem[{{Waxman}(2004)}]{Waxman2004}
{Waxman}, E. 2004, \apj, 602, 886

\bibitem[{{Woosley} \& {Bloom}(2006)}]{Woosley2006}
{Woosley}, S.~E., \& {Bloom}, J.~S. 2006, \araa, 44, 507

\bibitem[{{Yamazaki} {et~al.}(2003){Yamazaki}, {Yonetoku}, \&
  {Nakamura}}]{Yamazaki2003}
{Yamazaki}, R., {Yonetoku}, D., \& {Nakamura}, T. 2003, \apjl, 594, L79

\end{thebibliography}

\end{document}